\documentstyle[12pt,epsfig]{article}
\newcommand{\simlt}  {\raisebox{-.6ex}{$\stackrel{\textstyle <}{\sim}$}}
\newcommand{\simgt}  {\raisebox{-.6ex}{$\stackrel{\textstyle >}{\sim}$}}
\begin{document}                                                                
\begin{flushright}
RAL-TR-1999-026 \\
25 March 1999 \\
\end{flushright}                                                               
\begin{center}
{\Large  
A Redetermination of the
Neutrino Mass-Squared Difference
in Tri-Maximal Mixing            
with Terrestrial Matter Effects
}
\end{center}
\vspace{2mm}
\begin{center}
{P. F. Harrison\\
Physics Department, Queen Mary and Westfield College\\
Mile End Rd. London E1 4NS. UK \footnotemark[1]}
\end{center}
\begin{center}
{and}
\end{center}
\begin{center}
{D. H. Perkins\\
Nuclear Physics Laboratory, University of Oxford\\
Keble Road, Oxford OX1 3RH. UK \footnotemark[2]}
\end{center}
\begin{center}
{and}
\end{center}                      
\begin{center}                      
{W. G. Scott\\                      
Rutherford Appleton Laboratory\\    
Chilton, Didcot, Oxon OX11 0QX. UK \footnotemark[3]} 
\end{center}
\vspace{2mm}
\begin{abstract}
\baselineskip 0.6cm
\noindent 
We re-fit for the neutrino mass-squared difference 
$\Delta m^2$ in the threefold maximal 
(ie.\ tri-maximal) mixing scenario
using recent CHOOZ and SUPER-K data,
taking account of matter effects in the Earth.
While matter effects have
little influence on reactor experiments 
and proposed long-baseline accelerator experiments
with $L$ $\simlt$ $1000$ km,
they are highly significant for atmospheric experiments, 
suppressing naturally $\nu_e$ mixing and enhancing 
$\nu_{\mu} - \nu_{\tau}$ mixing,
so as to effectively remove the experimental distinction
between threefold maximal 
and twofold maximal $\nu_{\mu}-\nu_{\tau}$ mixing.
Threefold maximal mixing is fully consistent with
the CHOOZ and SUPER-K data and 
the best-fit value for the neutrino mass-squared difference is 
$\Delta m^2 \simeq (0.98 \pm
\raisebox{1.0ex}{0.30} \raisebox{-0.9ex}{\hspace{-6.7mm}0.23}
\hspace{1.5mm}) \times 10^{-3}$ eV$^2$.
\end{abstract}
\begin{center}
{\em To be published in Physics Letters B}
\end{center}
\footnotetext[1]{E-mail:p.f.harrison@qmw.ac.uk}
\footnotetext[2]{E-mail:d.perkins1@physics.oxford.ac.uk}
\footnotetext[3]{E-mail:w.g.scott@rl.ac.uk}
\newpage 
\baselineskip 0.6cm

\noindent {\bf 1. Introduction} \\

\noindent
New limits on neutrino oscillations
from the CHOOZ reactor experiment \cite{chz1} 
definitively rule out
threefold maximal lepton mixing with
$\Delta m^2 \sim 10^{-2}$ eV$^2$ \cite{hps1}.
In the meantime however,
the second-generation 
underground water-Cherenkov experiment
SUPER-KAMIOKANDE \cite{skam}
has superseded the 
older KAMIOKANDE experiment \cite{kam1}
as regards measurements of
the atmospheric neutrino anomaly,
with the latest experimental fits 
to $\nu_{\mu} - \nu_{\tau}$
mixing \cite{skos}
clearly favouring a much smaller
value of $\Delta m^2$ than
before \cite{kama}.
The aim of this paper
is to underline the fact \cite{ftmt}
that, 
in the light of such developments,
threefold maximal mixing is {\it per se}
far from excluded,
in particular we emphasise, 
when terrestrial matter effects \cite{pant}
are taken into account.

In this paper we re-fit for the
neutrino mass-squared difference $\Delta m^2$
in the threefold maximal mixing scenario
using the CHOOZ and SUPER-K data,
taking full account of 
terrestrial matter effects
which were hitherto neglected
\cite{hps1} \cite{hps3} \cite{wgs1}.
This simplest of all 
possible 
mixing schemes
then
continues to explain
the great majority 
of oscillation data
just as previously claimed,
but now with
$\Delta m^2 \sim 10^{-3}$ eV$^2$. \\

\noindent {\bf 2. The Vacuum Scenario} \\ 

\noindent
In vacuum 
in the threefold maximal mixing scenario,
the mixing matrix
relating 
the neutrino flavour eigenstates
$\nu_e$, $\nu_{\mu}$, $\nu_{\tau}$
to the neutrino mass eigenstates
$\nu_1$, $\nu_2$, $\nu_3$
(with masses $m_1 < m_2 < m_3$ respectively)
takes the symmetric or `democratic' form:
\begin{eqnarray}
     \matrix{  \hspace{0.1cm} \nu_1 \hspace{0.2cm}
               & \hspace{0.4cm} \nu_2 \hspace{0.2cm}
               & \hspace{0.4cm} \nu_3  \hspace{0.2cm} } 
                                      \hspace{0.4cm} \nonumber \\
\matrix{ e \hspace{0.2cm} \cr
         \mu \hspace{0.2cm} \cr
         \tau \hspace{0.2cm} }
\left( \matrix{ 1/3  &
                      1/3 &
                              1/3 \cr
                1/3 &
                    1/3 &
                             1/3  \cr
      \hspace{2mm} 1/3 \hspace{2mm} &
         \hspace{2mm}  1/3 \hspace{2mm} &
           \hspace{2mm} 1/3  \hspace{2mm} \cr } \right)
\end{eqnarray}
where here and throughout this paper
we give directly the {\em squares} of the
moduli of the mixing elements, in place of 
the mixing elements themselves.
The mixing phenomenology
is then completely determined
by the two independent
vacuum mass-squared differences
$\Delta m^2$ $>$ $\Delta m'^2$
(eg.\ $\Delta m^2 = m_3^2-m_2^2$, $\Delta m'^2 = m_2^2-m_1^2$). 
From the atmospheric data
we have
$\Delta m^2 \sim 10^{-3}$ eV$^2$
(see below) 
while from the solar data we found
$\Delta m'^2 < 0.9 \times 10^{-11}$ eV$^2$ \cite{hps1},
so that
in threefold maximal mixing,
two masses 
are effectively degenerate and 
the spectrum of mass-squared differences
is hierarchical. 

Neutrino oscillations
violate lepton flavour conservation
and as a function of
propagation length $L$
the matrix of normalised transition amplitudes
$A_{l'l}$ from a charged lepton state $l$ 
to a charged lepton state $l'$
is given directly by exponentiating 
the neutrino mass matrix squared 
$mm^{\dagger}$ ($\equiv m^2$)
in the flavour basis:
$A = \exp (-im^2L/2E)$,
where $E$ is the energy of the neutrino,
assumed to be relativistic.
The mixing matrix $U$
comprising the normalised eigenvectors of $m^2$
diagonalises the mass matrix thus:
$U^{\dagger}m^2U=$ diag $( \; m_1^2, \; m_2^2, \; m_3^2 \; )$.
In terms of the mixing matrix
the above matrix exponentiation is readily achieved
by first exponentiating the diagonal matrix of phases
and then rotating back to the flavour basis
(`{\em un}-diagonalising') as follows:
$A= U$ diag $( \; e^{i\phi_1}, \; e^{i\phi_2}, \; e^{i\phi_3} \; ) U^{\dagger}$,
where the phases $\phi_i = m_i^2L/2E$ for $i =1\dots 3$.
Thus any transition amplitude $A_{l'l}$
may be written as the sum
of three sub-amplitudes 
$A^{(i)}_{l'l}=X^{(i)}_{l'l} \; e^{i\phi_i}$, 
where
$X^{(i)}_{l'l}=U_{l'i}U^*_{li}$,
corresponding to the independent propagation
of each of three neutrino mass eigenstates 
as usual.
(Note that
hypothetical `sterile' neutrinos \cite{ster}
play no role in the threefold maximal mixing scenario 
and are not considered here.)

Survival and appearance probabilities
$P(l \rightarrow l) =|A_{ll}|^2$
and 
$P(l \rightarrow l')=|A_{l'l}|^2$
are then given by:
$P(l \rightarrow l)=|A_{ll}^{(1)}+A_{ll}^{(2)}+A_{ll}^{(3)}|^2$
and
$P(l \rightarrow l')=|A_{l'l}^{(1)}+A_{l'l}^{(2)}+A_{l'l}^{(3)}|^2$,
respectively.
In threefold maximal mixing 
the three sub-amplitudes
corresponding to the three mass eigenstates
are always of equal modulus $1/3$.
All vacuum survival probabilities $P(l \rightarrow l)$ 
and appearance probabilities 
$P(l \rightarrow l')$, $P(l' \rightarrow l)$ 
are independendent of flavour ($l,l'= e,\mu,\tau$).
In the limit $\Delta m'^2 \rightarrow 0$
only one relative phase need
be retained (eg.\ $\phi_3-\phi_2$, 
where $\phi_2 = \phi_1$ for $\Delta m'^2 = 0$) 
and we are led to:
\begin{equation}
P(l \rightarrow l) = 5/9 +4/9\cos(\Delta m^2L/2E) 
\end{equation}
and:
\begin{equation}
P(l \rightarrow l') = P(l' \rightarrow l) = 4/9-4/9\cos(\Delta m^2L/2E).
\end{equation}
Averaging over a range of $E$ (and/or $L$)
such that oscillating terms no longer contribute,
still in the limit $\Delta m'^2 \rightarrow 0$
with two sub-amplitudes adding `coherently',
one has for general mixing:
$<P(l \rightarrow l)> \; = \;
(X_{ll}^{(1)}+X_{ll}^{(2)})^2+(X_{ll}^{(3)})^2$
where the $X_{ll}^{(i)}$ are real 
(the matrices $X^{(i)}$ are hermitian), and:
$<P(l \rightarrow l')> \; = \;
|X_{l'l}^{(1)}+X_{l'l}^{(2)}|^2+|X_{l'l}^{(3)}|^2 = 2|X_{l'l}^{(3)}|^2$, 
where the last equality
relies on unitarity ($X^{1}+X^{2}+X^{3}=I$, the identity matrix).
In the threefold scenario,
one expects then a first `step' or `threshold'
at $L \sim (\Delta m^2/2E)^{-1}$,
marking the descent from 
$P(l \rightarrow l) =1$
to:
\begin{equation}
<P(l \rightarrow l)> \; =  (1/3+1/3)^2+(1/3)^2 = 5/9
\end{equation}
and the rise from
$P(l \rightarrow l') = P(l' \rightarrow l)= 0$
to:
\begin{equation}
<P(l \rightarrow l')> \; = \; <P(l' \rightarrow l)> \;
=2\times (1/3)(1/3)= 2/9.
\end{equation}
If $\Delta m'^2$ is non-zero 
there will be a second 
threshold at $L \sim (\Delta m'^2/2E)^{-1}$
beyond which 
all three sub-amplitudes 
add `incoherently' so that:
$<P( l \rightarrow l)> \; = \;
(X_{ll}^{(1)})^2+(X_{ll}^{(2)})^2+(X_{ll}^{(3)})^2$
with:
$<P( l \rightarrow l')> \; = \;
|X_{l'l}^{(1)}|^2+|X_{l'l}^{(2)}|^2+|X_{l'l}^{(3)}|^2$,
and in threefold maximal mixing:
$<P(l \rightarrow l)>\; = \; <P(l \rightarrow l')> \; 
= \; <P(l' \rightarrow l)> \; = 1/3$.

For solar neutrinos,
in the threefold maximal mixing scenario, 
matter effects in the Sun are expected
to be rather unimportant \cite{hps2}
and the vacuum prediction
Eq.~(4) may be compared directly
with the measured suppressions,
at least
for the gallium experiments.
The same comparison for the
SUPER-K \cite{sksn} and HOMESTAKE \cite{home} solar results,
(correcting for the neutral current contribution
to the $\nu e \rightarrow \nu e$ rate in SUPER-K)
is limited
by {\it de facto} uncertainties
in the $^8$B flux \cite{hps3}.

In the atmospheric experiments
the initial beam comprises both $\nu_{\mu}$ and $\nu_e$,
and account must be taken of
$\nu_{\mu} \leftrightarrow \nu_e$
transitions, 
which tend to compensate 
the survival rates in Eq.~(4).
In the approximation that
the flux ratio 
at production
$\phi(\nu_{\mu})/\phi(\nu_e) =2/1$
the effective $\nu_{\mu}$ 
suppression in vacuum becomes:
\begin{equation}
5/9 + 1/2 \times 2/9 = 2/3
\end{equation}
while the effective $\nu_e$ suppression becomes:
\begin{equation}
5/9+ 2 \times 2/9 = 1
\end{equation}
so that the $\nu_e$ rate
is perfectly compensated
in this case.
The initial flux ratio
is indeed $\phi(\nu_{\mu})/\phi(\nu_e) \sim 2/1$
for $E$ $\simlt$ $1$ GeV
(increasing
with $E$ thereafter).

While this concludes our
discussion of the vacuum scenario,
as we turn to consider
the influence of matter effects
it will prove useful
to denote vacuum quantities
with argument $(0)$ for vacuum explicitly,
thus we define: $m^2(0) \equiv m^2$, 
$U(0) \equiv U$,
$\nu_i(0) \equiv \nu_i$, 
$m_i(0) \equiv m_i$,
$\Delta m^2(0) \equiv \Delta m^2$,
$\Delta m'^2(0) \equiv \Delta m'^2$ etc. \\
 
\noindent {\bf 3. Terrestrial Matter Effects} \\

\noindent
In any scheme 
with $\nu_e$ mixing (eg.\ Eq.~1),
matter effects in the Earth,
in particular
for atmospheric neutrinos, 
can be very significant indeed.
Matter effects result from 
the forward scattering
of $\nu_e$ from electrons in the matter,
modifying the mass matrix by
adding a term 
to the $(e,e)$ entry
of the vacuum mass matrix,
proportional to 
the matter density $\rho$.
Basic trends
in matter phenomenology
can 
usefully be anticipated
by noting that
in the limit
of infinite density ($\rho \rightarrow \infty$),
the $\nu_e$ becomes
the heaviest neutrino,
which being then itself
a mass eigenstate,
completely decouples
in the mixing.
In general clearly,
before this limit is approached,
masses, mixings etc.\
are functions of the
matter density, 
hence we have: $m^2(\rho)$, $U(\rho)$,
$\nu_i(\rho)$, $m_i(\rho)$,
$\Delta m^2(\rho)$,
$\Delta m'^2(\rho)$ etc.,
or more directly
functions of the number density 
$N_e$ of electrons in the matter.

In the present analysis  
matter effects are incorporated
using the general $3 \times 3$
numerical program described 
in Ref.\ \cite{hps2},
where for a hierarchical spectrum
$\Delta m^2(0) \gg \Delta m'^2(0)$,
the formulae of 
Ref.~\cite{gnmt} apply.
Our numerical calculation proceeds in steps 
$\Delta L = 100$ km through the Earth,
with the density and composition of
the Earth input as a function of depth
from a recent tabulation \cite{jaco}.
Before giving 
our full numerical results 
in any detail however,
in this section 
we sketch briefly
the main trends predicted analytically
in threefold maximal mixing
with $\Delta m'^2(0) \ll \Delta m^2(0)$,
as a function of increasing 
matter density,
or equivalently as a function of increasing $E$.
 
We identify
three significantly distinct 
density regimes
where the mixing phenomenolgy
remains essentially constant
over a wide range of scales.

Firstly, clearly
at very low matter densities
$\sqrt{2}GN_e \ll \Delta m'^2(0)/E$
(where $G$ is the Fermi constant)
vacuum mixing remains essentially valid.
Conceivably if
$\Delta m'^2(0)$ is extremely small
or zero ($\Delta m'^2$ $\simlt$ $10^{-30}$ eV$^2$) this regime
may not be realised in nature,
even for neutrinos from distant supernovae
propagating in deep space 
($N_e \sim 10^3$ m$^{-3}$).

Next, at `intermediate' densities
ie.\ for $\Delta m'^2(0)/E \ll \sqrt{2}GN_e \ll \Delta m^2(0)/E$,
matter effects lift the effective 
degeneracy between the two light neutrinos,
such that in matter for neutrinos: 
$\nu_1(\rho) \rightarrow [\nu_1(0) - \nu_2(0)]/\sqrt{2}$ and
$\nu_2(\rho) \rightarrow [\nu_1(0) + \nu_2(0)]/\sqrt{2}$
(up to phases),
where the lighter matter mass eigenstate $\nu_1(\rho)$
has zero $\nu_e$ content.
The mixing matrix is thereby
deformed in matter as follows:
\begin{eqnarray}
     \matrix{  \hspace{0.06cm} \nu_1(0) \hspace{0.06cm}
               & \hspace{0.06cm} \nu_2(0) \hspace{0.06cm}
               & \hspace{0.06cm} \nu_3(0)  \hspace{0.06cm} }
\hspace{3.3cm}
      \matrix{  \hspace{0.000cm} \nu_1(\rho) \hspace{0.000cm}
               & \hspace{0.000cm} \nu_2(\rho) \hspace{0.000cm}
               & \hspace{0.000cm} \nu_3(\rho)  \hspace{0.000cm} } 
                                             \hspace{0.5cm} \nonumber \\
\matrix{ e \hspace{0.2cm} \cr
         \mu \hspace{0.2cm} \cr
         \tau \hspace{0.2cm} }
\left( \matrix{ 1/3  &
                      1/3 &
                              1/3 \cr
                1/3 &
                    1/3 &
                             1/3  \cr
      \hspace{2mm} 1/3 \hspace{2mm} &
         \hspace{2mm}  1/3 \hspace{2mm} &
           \hspace{2mm} 1/3  \hspace{2mm} \cr } \right)
\hspace{0.5cm} \longrightarrow \hspace{0.5cm}
\matrix{ e \hspace{0.2cm} \cr
         \mu \hspace{0.2cm} \cr
         \tau \hspace{0.2cm} }
\left( \matrix{ .  &
                      2/3 &
                              1/3 \cr
                1/2 &
                    1/6 &
                             1/3  \cr
      \hspace{2mm} 1/2 \hspace{2mm} &
         \hspace{2mm}  1/6 \hspace{2mm} &
           \hspace{2mm} 1/3 \hspace{2mm} \cr } \right)
\end{eqnarray}
with the smaller mass-squared difference
$\Delta m'^2(\rho)/2E \rightarrow 2/3 \times \sqrt{2}GN_e$
in this case.
The factor $2/3$ multiplying
the matter mass scale
reflects the fact that 
only two out of the three
relevant
$2 \times 2$ sub-determinants
of the vacuum mass matrix 
are modified by the matter term. 
(Symmetrically,
this same factor of $2/3$ 
will appear again
multiplying the vacuum mass scale
in the high density limit, see below).
In the Earth's mantle ($\rho \sim 5$ g cm$^{-3}$)
the matter mass-scale \cite{wolf} 
is $\sqrt{2}GN_e \simeq 0.38 \times 10^{-3}$ eV$^2$/GeV,
and the corresponding length-scale is
$(\sqrt{2}GN_e)^{-1} \simeq 1040$ km/rdn.
Thus 
the limit
Eq.~(8) should
be physically relevant
for atmospheric neutrinos,
assuming $\Delta m^2(0) \sim 10^{-3}$ eV$^2$,
for neutrino energies
$E$ $\simlt$ $1$ GeV.
 
The matrix 
on the right-hand-side of Eq.~(8)
is one of the matrices with the
$\nu_3$ maximally mixed 
introduced in Ref.~\cite{hps3}.
For $L$ $\simlt$ $(\sqrt{2}GN_e)^{-1}$
the vacuum mixing predictions Eqs.~(2)-(7)
are reproduced.
For $L$ $\simgt$ $(\sqrt{2}GN_e)^{-1}$,
ie.\ beyond the `matter threshold',
the $\nu_{\mu}$ survival probability
averaged over $E$ and $L$,
is from row 2 of Eq.~(8):
\begin{equation}
<P(\mu \rightarrow \mu)> \; = (1/2)^2 + (1/6)^2 + (1/3)^2 = 7/18
\end{equation}
while the average $\nu_e$ survival probability
is (from row 1):
\begin{equation}
<P(e \rightarrow e)> \; = (0)^2 + (2/3)^2 + (1/3)^2 =5/9
\end{equation}
just as it was
after the first threshold, defined by $\Delta m^2$,
in the vacuum analysis above.
Likewise the corresponding 
$\nu_{\mu} \leftrightarrow \nu_e$
appearance probabilities are:
\begin{equation}
<P(\mu \rightarrow e)> \; = \; <P(e \rightarrow \mu)>\; 
= (0)(1/2) + (2/3)(1/6)+(1/3)(1/3) = 2/9
\end{equation}
(from rows 1 and 2) again as in the vacuum analysis.

For atmospheric neutrinos,
in the approximation
that the initial flux ratio
$\phi(\nu_{\mu})$ $/$ $\phi(\nu_e)$ $=$ $2/1$,
the effective $\nu_{\mu}$ suppression becomes:
\begin{equation} 
7/18 + 1/2 \times 2/9 = 1/2
\end{equation}
replacing the vacuum 
prediction of $2/3$ from Eq.~(6) \cite{hps1}\cite{hps3}\cite{wgs1}, 
for $L \gg (\sqrt{2}GN_e)^{-1}$.
The $\nu_e$ rate 
however remains
perfectly compensated
just as it was for $\phi(\nu_{\mu})/\phi(\nu_e)$ $=$ $2/1$
in the vacuum analysis and Eq.~(7) applies.
Note that these
final suppression factors 
are exactly the same as would be 
expected in twofold maximal
$\nu_{\mu}-\nu_{\tau}$ mixing.

Finally, for sufficiently high matter densities
$\Delta m^2(0)/E \ll \sqrt{2}GN_e$
(or equivalently for sufficiently high neutrino energies),
threefold maximal mixing tends physically to
twofold maximal $\nu_{\mu} - \nu_{\tau}$ mixing
with $\Delta m^2(0)$ fixing
the smaller mass-squared difference 
$\Delta m'^2(\infty) \rightarrow 2/3 \times \Delta m^2(0)$ in matter,
and with the larger matter mass-squared difference
fixed by the matter mass-scale
$\Delta m^2(\infty)/2E \rightarrow \sqrt{2}GN_e$:
\begin{eqnarray}
     \matrix{  \hspace{0.0cm} \nu_1(\rho) \hspace{0.0cm}
               & \hspace{0.0cm} \nu_2(\rho) \hspace{0.0cm}
               & \hspace{0.0cm} \nu_3(\rho)  \hspace{0.0cm} }
\hspace{3.4cm}
     \matrix{  \hspace{0.0cm} \nu_1(\infty) \hspace{0.0cm}
               & \hspace{0.0cm} \nu_2(\infty) \hspace{0.0cm}
               & \hspace{0.0cm} \nu_3(\infty)  \hspace{0.0cm} } 
                                   \hspace{0.3cm} \nonumber \\
\matrix{ e \hspace{0.2cm} \cr
         \mu \hspace{0.2cm} \cr
         \tau \hspace{0.2cm} }
\left( \matrix{ .  &
                      2/3 &
                              1/3 \cr
                1/2 &
                    1/6 &
                             1/3  \cr
      \hspace{2mm} 1/2 \hspace{2mm} &
         \hspace{2mm}  1/6 \hspace{2mm} &
           \hspace{2mm} 1/3  \hspace{2mm} \cr } \right)
\hspace{0.5cm} \longrightarrow \hspace{0.5cm}
\matrix{ e \hspace{0.2cm} \cr
         \mu \hspace{0.2cm} \cr
         \tau \hspace{0.2cm} }
\left( \matrix{ .  &
                      . &
                              1 \cr
                1/2 &
                    1/2 &
                             .  \cr
      \hspace{3.5mm} 1/2 \hspace{3.5mm} &
         \hspace{3.5mm}  1/2 \hspace{3.5mm} &
           \hspace{3.5mm} . \hspace{3.5mm} \cr } \right) .
\end{eqnarray}
As foreseen,
with the matter interaction
affecting $\nu_e$ directly,
$\nu_e \rightarrow \nu_3$
asymptotically. 
Thus the higher frequency
oscillations have zero amplitude
whereby, for example, 
the $\nu_{\tau}$ appearance probability 
in a high energy $\nu_{\mu}$ beam becomes simply:
\begin{equation}
P(\mu \rightarrow \tau)
= 1/2 -1/2 \cos[2/3 \times \Delta m^2(0)L/2E] .
\end{equation}
For $L \rightarrow 0$ vacuum predictions
must always be reproduced \cite{pant}.
Clearly
Eq.~(14) coincides with Eq.~(3)
for $L \rightarrow 0$.
For $L \rightarrow \infty$ 
we have
$<P(\mu \rightarrow \tau)> \; \rightarrow 1/2$,
so that $\nu_{\tau}$ appearance
may be said to be 
enhanced by matter effects, cf.\ Eq.~(5),
and correspondingly
$<P(\mu \rightarrow \mu)> \; \rightarrow 1/2$
in that case.
As the $\nu_e$ 
decouples completely
$P(e \rightarrow e) \rightarrow 1$
with $P(\mu \rightarrow e)= P(e \rightarrow \mu) \rightarrow 0$
in the limit,
so that
in the atmospheric experiments
there are 
no significant compensation effects
expected at the highest energies,
despite the fact that
the production flux ratio 
$\phi(\nu_{\mu})/\phi(\nu_e)$
is becoming large,
$\phi(\nu_{\mu})/\phi(\nu_e)$ $\simgt$ $3/1$
for $E$ $\simgt$ $10$ GeV.

The important conclusion
is that threefold maximal mixing
with terrestrial matter effects 
{\em exactly} mimics 
twofold $\nu_{\mu}-\nu_{\tau}$ mixing
for the atmospheric neutrino rates
both at the lowest energies
and at the highest energies,
but for somewhat different reasons
in the two cases, as we have seen.
Observable differences
between threefold and twofold mixing
are expected at best only in a window
of `intermediate' energies,
where 
$\phi(\nu_{\mu})/\phi(\nu_e)$ $>$ $2/1$
but before the limit Eq.~(13)
is approached.
Of course our numerical results
reported below
incorporate the full expected
energy and zenith-angle
dependence of the
flux ratio at production 
$\phi(\nu_{\mu})/\phi(\nu_e)$ 
\cite{dhp1},
with appropriate averaging
over detailed 
$\nu/\bar{\nu}$ differences, 
neglected in the above discussion. \\

\noindent {\bf 4. Multi-GeV Zenith-Angle Distributions} \\

\noindent
When oscillations
are not individually resolved
the neutrino mass-squared difference
is determined by the
location of the corresponding threshold
on the $L/E$ scale
(the location of the `matter threshold' 
is of course energy-independent,
and is arguably better studied as a function
of $L$ rather than $L/E$).
For atmospheric neutrinos
the neutrino flight-path $L$ is related to 
the zenith angle $\theta$ of the neutrino by
\begin{equation}
L=\sqrt{R_{\oplus}^2\cos^2\theta+2R_{\oplus}H+H^2}-R_{\oplus}\cos\theta
\end{equation}
where $R_{\oplus} \simeq 6380$ km is the radius of the Earth 
and $H \sim 20$ km is the effective height of the atmosphere.
In the water-Cherenkov experiments
the neutrino zenith angle
is estimated using the 
measurement of the outgoing charged-lepton,
which will best correlate 
with the neutrino direction
at higher energies,
so that we expect the most
reliable information
to come from the so-called multi-GeV sample
(ie.\ events with charged lepton momentum 
$p$ $\simgt$ $1.3$ GeV for electrons 
and $p$ $\simgt$ $1.4$ GeV for muons).

Fig.~1 shows the measured
zenith angle distributions
for multi-GeV events,
combining the SUPER-K \cite{skos} and KAMIOKA \cite{kama} data
for maximum statististical weight.
There is clear evidence 
for a step (or `threshold') 
as a function of $\cos \theta$
with an approximate $50\%$ suppression
of the $\mu$-like events
for $\cos \theta < -0.2$.
At the same time
no such effect is apparent in the
corresponding distribution 
for $e$-like events, Fig.~1b.
Since a full monte-carlo simulation
including detector effects etc.\
is beyond the scope of the present paper,
we have
wherever possible
in this analysis
made appropriate use
of the expected event rates
for no oscillations
given by the experimenters themselves.
Thus in Fig.~1
we plot the ratio of
observed to expected events
in preference to the event rates themselves,
in order to minimise dependence
on detector acceptance etc.
The plotted ratio
in each case
is normalised to the
threefold maximal prediction
with matter effects for
$\Delta m^2 = 0.98 \times 10^{-3}$ eV$^2$
(which is the overall
best-fit $\Delta m^2$ 
in threefold maximal mixing, 
see Section~5 below)
so that it is only the shape of
the zenith-angle dependence
which is being tested here,
with no reliance
on absolute fluxes.

In Fig.~1 the solid curves 
represent the threefold maximal mixing predictions
computed taking full account of terrestrial matter effects
and compensation effects
for $\Delta m^2 = 0.98 \times 10^{-3}$ eV$^2$,
with the dotted curves showing the 
corresponding vacuum predictions.
The dashed curves represent
twofold maximal $\nu_{\mu} - \nu_{\tau}$ mixing
for the same value of $\Delta m^2$.
All the above curves 
are averaged over the multi-GeV energy distribution
and incorporate angular smearing.
We take gaussian smearing
in angle around the neutrino direction
with a width fixed from
existing $\nu/\bar{\nu}$ data \cite{phd1},
where the width 
falls with energy
proportional to $1/\sqrt{E}$.
The mean angle between the
neutrino direction and the charged-lepton direction
is then $21^o$ for the multi-GeV sample
(and $43^o$ for subGeV events
with $p > 400$ MeV/c, see Section~5 below).
For comparison, angular smearing 
resulting from the Cherenkov method itself
has the same energy dependence,
but
is an order of magnitude smaller \cite{naka}
and may be neglected.
Particle mis-identification effects,
ie.\ mis-classification of
$\mu$-events as 
$e$-events and vice versa,
believed to become important
only at the percent level \cite{evmu},
are also neglected in our analysis.
As is clear from Fig.~1
(and see also Table~1, Section~5 below)
both threefold maximal mixing with matter effects 
and twofold maximal
$\nu_{\mu}-\nu_{\tau}$
mixing 
are consistent with 
the measured multi-GeV 
zenith-angle distributions.

The predicted up/down ratios $(U/D)_{\mu}$ and $(U/D)_e$
for multi-GeV $\mu$-like and $e$-like events respectively,
are shown in Fig.~2,
as functions of $\Delta m^2$,
for each of the various mixing scenarios above.
The up/down ratios are
defined from Fig.~1
as the ratio of up to down
rates 
with $|\cos \theta | > 0.2$
and are expected \cite{ftud}
to be relatively free
of systematic effects,
eg.\ flux uncertainties.
At multi-GeV energies
the initial flux ratio 
$\phi(\nu_{\mu})/\phi(\nu_e) > 2/1$,
and in the case of threefold maximal mixing 
one expects to see
{\it over}-compensation of the $\nu_e$ rate, 
ie.\ $(U/D)_e > 1$, cf.\ Eq.~(7),
and {\it under}-compensation of the $\nu_{\mu}$ rate, 
ie.\ $(U/D)_{\mu}$ $<$ $1/2$, cf.\ Eq.~(12)
(with smearing $(U/D)_{\mu}$ $\simeq$ $1/2$, Fig.~2),
at least 
for $\Delta m^2$ $\simgt$ $10^{-3}$ eV$^2$.
As is clear from Fig.~2, 
for $\Delta m^2$ $\simlt$ $10^{-3}$ eV$^2$
over-compensation of the $\nu_e$ rate
is suppressed relative to the vacuum prediction
as a result of terrestrial matter effects,
with $(U/D)_e$ approaching unity
as $\Delta m^2 \rightarrow 0$,
simulating
$\nu_{\mu}-\nu_{\tau}$ mixing as discussed above.

In Fig.~2 the data points with error bars
represent the measured up/down ratios
based on the combined SUPER-K and KAMIOKA data:
$(U/D)_{\mu}=0.53 \pm$  
\raisebox{0.9ex}{0.05} \raisebox{-0.9ex}{\hspace{-8.9mm}0.04}
and $(U/D)_e = 0.99 \pm$ 
\raisebox{0.9ex}{0.11} \raisebox{-0.9ex}{\hspace{-8.9mm}0.10},
these being plotted arbitrarily at 
$\Delta m^2 \sim 0.98 \times 10^{-3}$ eV$^2$,
and extended by the shaded bands.
The up/down ratio for all events
$(U/D)_{e+\mu}=0.68 \pm 0.05$ (not shown)
is less incisive, 
but has the advantage of being
independent of particle misidentifiaction effects.
Clearly the data
on the up/down ratios
are consistent with
either twofold maximal $\nu_{\mu}-\nu_{\tau}$ mixing,
or with threefold maximal mixing with
inclusion of terrestrial matter effects,
for $\Delta m^2 \sim 10^{-3}$ eV$^2$. \\

\noindent{\bf 5. The Sub-GeV Data, R-values and Upward Muons} \\

\noindent
The KAMIOKA/SUPER-K
collaborations
have also given
zenith angle distributions
for sub-GeV events
(lepton momentum $p < 1.3$ GeV),
as well as values for the double ratio
$R = (\mu/e)_{DATA}/(\mu/e)_{MC}$
for sub-GeV and multi-GeV events,
and independent data
on upward muons.
For some of these data-sets
there are
significant systematic effects
to be considered, against
the gain in statistiscal weight.

Fig.~3 shows the
observed zenith angle dependence
for the higher energy 
($p > 400$ MeV) subset 
of the sub-GeV sample 
in the SUPER-K experiment.
The dashed curve
is the prediction of threefold maximal mixing
for $\Delta m^2 \simeq 0.98 \times 10^{-3}$ eV$^2$
including terrestrial matter effects
and compensation effects,
showing clearly the `matter threshold' 
and the matter-induced oscillations,
but neglecting angular smearing.
The solid curve 
includes the expected 
effect of angular smearing
as discussed previously,
and which is now very significant.
A fit to the data of Fig.~3
in threefold or twofold mixing 
yields a best-fit
$\Delta m^2 \sim 2 \times 10^{-3}$ eV$^2$
($\chi^2$/DOF $\simeq 5.0/7$, CL$ \simeq 66$\%).
The zenith-angle data 
for the lower energy ($p < 400$ MeV)
subset of the sub-GeV data
are not included in this analysis.
Angular smearing
and geomagnetic effects \cite{bart} 
are expected to dominate
the zenith-angle dependence
at such low energies, where
very little zenith-angle dependence 
is seen and essentially
no $\Delta m^2$ information survives.

The integrated ratio of ratios
$R = (\mu/e)_{DATA}/(\mu/e)_{MC}$
is independent of smearing effects,
but dependent on flux uncertainties
which affect the flavour ratio 
$\phi(\nu_{\mu})/\phi(\nu_e)$,
particularly since
$e$-like and $\mu$-like events have different
Cherenkov thresholds.
The latest $R$-values given
by the SUPER-K experiment are:
$R = 0.67 \pm 0.02 \pm 0.05$
for the sub-GeV sample and
$R = 0.66 \pm 0.04 \pm 0.08$
for the multi-GeV sample,
where the first error is statistical
and the second is the systematic error
in each case, 
as given by SUPER-K.
These results
are shown in Fig.~4
by the data points and shaded bands,
together with the predictions
from threefold maximal mixing (solid curves) 
and twofold $\nu_{\mu}-\nu_{\tau}$ mixing (dashed curves),
plotted as a function of $\Delta m^2$.
Clearly within the 
combined statistical and systematic errors 
the consistency with
$\Delta m^2 \sim 10^{-3}$ eV$^2$
is perfectly acceptable.
It is perhaps worth
commenting here
that threefold maximal mixing
goes some way to
resolving the problem
of the near-equality
of the multi-GeV and sub-GeV
$R$-values
raised by LoSecco \cite{lsco}.

In the underground detectors
upward muons result from neutrino interactions
in the rock surrounding the detector 
\cite{maco} \cite{kaum}.
Fig.~5 shows the double ratio $S$
of stopping to through muons
for data divided by monte-carlo:
$S=0.59 \pm 0.06 \pm 0.08$
as measured by SUPER-K \cite{skum}
(data point and shaded band).
The solid curve is our threefold maximal mixing
prediction with terrestrial matter effects included,
plotted as a function of $\Delta m^2$,
while the dashed curve 
corresponds to twofold maximal
$\nu_{\mu}-\nu_{\tau}$ mixing.
The shift of a factor of $2/3$ 
in $\Delta m^2$
between the two curves 
as expected at high energies, 
see Eq.~(14), is apparent,
with threefold maximal mixing
showing a slightly lower minimum
value for $S$,
due to terrestrial matter effects.
The relatively low measured value of $S$
seems to point here 
to $\Delta m^2 \sim 10^{-3}$ eV$^2$
in either threefold or twofold mixing. \\
 
\noindent {\bf 6. The CHOOZ Data and Overall Fit} \\

\noindent
The CHOOZ experiment \cite{chz1}
measures the survival probability
$P(\bar{e} \rightarrow \bar{e})$
for $\bar{\nu_e}$ by comparing
the observed to expected rates
for the reaction
$\bar{\nu}_e + p \rightarrow e^+ +n$
at a distance $L \sim 1$ km
from the CHOOZ reactor site.
The initial result
for $P(\bar{e} \rightarrow \bar{e})$
is shown in Fig.~6,
plotted as a function
of antineutrino energy $E$,
related to the 
measured positron energy $E_e$
by $E = 1.8 + E_e$ MeV.
The data so far are consistent
with $P(\bar{e} \rightarrow \bar{e}) =1$,
and no oscillation signal is claimed.

At such short pathlengths 
$L \ll (\sqrt{2}GN_e)^{-1}$, 
matter effects can be
safely neglected,
and in Fig.~6 the solid curve
is the threefold maximal mixing prediction
calculated from 
the vacuum formula Eq.~(2)
for $\Delta m^2 = 0.98 \times 10^{-3}$ eV$^2$.
Twofold maximal $\nu_{\mu}-\nu_{\tau}$ mixing
predicts $P(\bar{e} \rightarrow \bar{e})=1$
as shown by the dashed line,
independent of matter effects
and independent of $E$ (and $\Delta m^2$).
The best-fit to the data of Fig.~6
in threefold maximal mixing is:
$\Delta m^2 = 0.60 \times 10^{-3}$ eV$^2$
($\chi^2$/DOF $=10.6/9$, CL$=30$\%),
although clearly the data
are also fully consistent with
$\Delta m^2 \rightarrow 0$,
($\chi^2$/DOF $\rightarrow 11.0/9$, CL$=28$\%),
or equivalently with 
$\nu_{\mu}-\nu_{\tau}$ mixing.

Our overall fit
for the neutrino mass-squared difference $\Delta m^2$
is based on the atmospheric data
already discussed,
taken together with 
the CHOOZ data
for $P(\bar{e} \rightarrow \bar{e})$
as a function of $E$.
Specifically we compute the total $\chi^2$
summed over all the bins
displayed in  
Figs.~1,~3,~4,~5,~6
as a function of $\Delta m^2$
for twofold maximal $\nu_{\mu}-\nu_{\tau}$ mixing
and for threefold maximal mixing
with matter effects.
There are 33 data points 
in total,
but 4 normalisation
factors to be determined in each case
(one for each zenith-angle plot)
in addition to $\Delta m^2$,
so that there are $33-5=28$
DOF
in each fit. 

The results
of the overall fits 
are shown in Fig.~7.
For twofold maximal 
$\nu_{\mu}-\nu_{\tau}$ mixing
(dashed curve)
there is a broad minimum in $\chi^2$
extending over the range
$\Delta m^2 \sim 7 \times 10^{-4} - 4 \times 10^{-3}$ eV$^2$
with the best-fit
$\Delta m^2 \sim 2.2 \times 10^{-3}$ eV$^2$
corresponding to an excellent fit
($\chi^2/$DOF $= 18.7/28$, CL $=91$\%).
For threefold maximal mixing
(solid curve)
the mass-squared difference 
is highly constrained
by the CHOOZ data 
giving a much narrower
minimum in $\chi^2$, 
and a significantly
more restricted $\Delta m^2$-range:
$\Delta m^2 \simeq (0.98 \pm 
\raisebox{0.9ex}{0.30} \raisebox{-0.9ex}{\hspace{-7.3mm}0.23}
\hspace{0.4mm}) \times 10^{-3}$ eV$^2$,
however still corresponding to a very good fit
($\chi^2/$DOF $= 25.4/28$, CL $=61$\%).
A full breakdown
of the $\chi^2$-contributions
coming from
the various data-sets
is given in Table~1.

While both 
threefold maximal mixing
and
twofold maximal 
$\nu_{\mu}-\nu_{\tau}$ mixing
are consistent with the
combined data then,
it is clear that
in threefold maximal mixing
the best-fit value of $\Delta m^2$
is extremely close to the 
published CHOOZ limit \cite{chz1}
on $\Delta m^2$ computed for the case of
twofold maximal $\nu_{\mu}-\nu_e$ mixing.
This near-coincidence
of the best-fit value
with the published upper-limit 
strongly suggests that
if threefold maximal mixing is valid, 
either of the 
two existing long-baseline reactor experiments
CHOOZ \cite{chz1} or PALO-VERDE \cite{palo}
could,
with continued running, 
increased detector mass
and especially with increased
baseline $L \sim 5-10$ km,
still be the first experiments to observe
`man-made' neutrino oscillations,
and to establish the existence of $\nu_e$ mixing
at the `atmospheric scale' 
$\Delta m^2 \sim 10^{-3}$ eV$^2$.
Otherwise the KAMLAND experiment \cite{kmln}
should be decisive. \\

\noindent{\bf 7. Perspective and Future Prospects} \\

\noindent
Naturally there are
many other mixing schemes which can
similarly describe the data.
We mention explicitly here the Fritzsch-Xing hypothesis \cite{xing},
and in particular also the `bi-maximal' hypothesis \cite{bimx}
and its generalisations \cite{alt1} \cite{jarl}
in which the $\nu_3$ is assumed to have
$\nu_e$, $\nu_{\mu}$, $\nu_{\tau}$
content $ \; 0$, $1/2$, $1/2 \;$
perfectly replicating
the phenomenology of twofold
$\nu_{\mu}-\nu_{\tau}$ mixing 
as far as the atmospheric
experiments are concerned.
In such schemes
the zero or near-zero in the
top right-hand corner of the mixing matrix 
guarantees no $\nu_e$ mixing at the
atmospheric scale $\Delta m^2 \sim 10^{-3}$ eV$^2$.

Clearly certain `bi-maximal' schemes
can mimic `tri-maximal' mixing
(ie.\ threefold maximal mixing)
extremely effectively.
For example a bi-maximal scheme
with the $\nu_2$ (tri-) maximally mixed
(ie.\ having 
$\nu_e$, $\nu_{\mu}$, $\nu_{\tau}$
content 
$1/3$, $1/3$, $1/3$),
would have very similar
phenomenology to threefold maximal mixing
even for the solar data,
with
$P(e \rightarrow e) =5/9$
in the gallium experiments
and
with the added possibility
to exploit a large angle MSW
solution with
$P(e \rightarrow e) =1/3$
in the `bathtub',
in the higher energy experiments,
for some particular
$\Delta m'^2 \sim 10^{-5}$ eV$^2$.
We emphasise however that
in the threefold maximal mixing scenario
suppression of $\nu_e$ mixing at the
atmospheric scale $\Delta m^2 \sim 10^{-3}$ eV$^2$,
occurs {\em naturally}
as a result of terrestrial matter effects
without the insertion of an arbitrary zero
in the mixing matrix
as in the bi-maximal schemes.

As regards `bi-maximal' 
versus `tri-maximal' mixing then,
the crucial experimental question
would seem to be whether or not there
is appreciable vacuum $\nu_e$ mixing 
at the atmospheric scale.
This question might perhaps be answered
in the future by the SUPER-K experiment itself,
with the observation of a significant
upward excess of $e$-like events
in the atmospheric data
at multi-GeV energies.
The predicted $U/D$ ratio
for $\nu_e$
in threefold maximal mixing,
when matter effects are included,
peaks 
as a function of neutrino energy
at $E \sim 7$ GeV,
where $(U/D)_e \simeq 1.13$.

Together with reactor experiments,
long-baseline accelerator experiments
using high-energy $\nu_{\mu}$-beams
will also probe $\nu_e$ mixing.
Fig.~8 shows the predicted 
energy-averaged
appearance probability $<P(\mu \rightarrow e)>$
for various mean neutrino energies
in threefold maximal mixing
for $\Delta m^2 = 1.0 \times 10^{-3}$ eV$^2$.
The curves are calculated
with and without matter effects
(solid and dotted curves respectively)
and for visual clarity
are shown averaged ($50:50$) over
$\nu/\bar{\nu}$ beams.
The suppression of $\nu_e$ appearance
as a result of matter effects
is plainly visible at large $L$,
but for all proposed experiments
with $L$ $\simlt$ $1000$ km,
matter effects can be
largely neglected. \\ 

\noindent{\bf 8. Conclusion} \\

In this paper we have shown that the recent data
from CHOOZ and SUPER-K are
consistent with threefold maximal mixing
(ie.\ `tri-maximal' mixing) 
for a neutrino mass-squared difference 
$\Delta m^2 \simeq (0.98 \pm 
\raisebox{0.9ex}{0.30} \raisebox{-0.9ex}{\hspace{-7.3mm}0.23}
\hspace{0.4mm}) \times 10^{-3}$ eV$^2$.
With just a very few acknowledged exceptions,
notably the LSND appearance result \cite{lsnd} 
and possibly 
also the HOMESTAKE solar measurement \cite{home},
the totality of previous data 
are also known \cite{hps3} to be in good agreement 
with this hypothesis \cite{hps1}.

Assuming tri-maximal mixing therefore,
and taking a hierarchical spectrum of neutrino
masses similar to that of the quarks and leptons,
we have for the heavy neutrino mass $m_3 \sim 30 \pm 5$ meV.
This result is consistent with a `see-saw' relation
to the up-type quarks for a heavy 
(right-handed) Majorana mass 
$M_R \sim 1.0 \times 10^{15}$ GeV.
The compton wavelenth of the heavy neutrino
$\nu_3$ is $\lambda_3 \sim 1/20$ mm.
Such a neutrino
would make little contribution
to cosmological dark matter.  

\vspace{5mm}
\noindent {\bf Aknowledgement}

\noindent It is a pleasure to thank R.~Edgecock, R.~Foot, E.~Kearns,
E.~Lisi, M.~Messier and D.~Wark 
for helpful comments/discussions/correspondence.

\newpage

.
\vspace{2.0cm}
\begin{center}
\begin{tabular}{|c|c|c|}
\hline
\hline
              &     &     \\
    & \hspace{4mm} 3-Fold Max.\ (mat.) \hspace{4mm} &
        \hspace{3mm} 2-Fold Max. ($\nu_{\mu}-\nu_{\tau}$) \hspace{3mm} \\
              & $\Delta m^2 = 0.98 \times 10^{-3}$ eV$^2$ &
                        $\Delta m^2 = 2.2 \times 10^{-3}$ eV$^2$  \\
              &     &     \\
\hline
\hline
              &     &     \\
  \hspace{3mm} Zenith-Angle \hspace{3mm}
                    & $\chi^2$/DOF $=$ $6.5/7$ &
                                        $\chi^2$/DOF $=$ $2.0/7$ \\
  Multi-GeV      & (CL $=$ $48$\%) &            (CL $=$ $96$\%)    \\
              &     &     \\
\hline
    +         &     &     \\
 Zenith-Angle  & $\chi^2$/DOF $=$ $11.9/15$ & $\chi^2$/DOF $=$ $7.1/15$ \\
   Sub-GeV      & (CL $=$ $69$\%) & (CL $=$ $95$\%) \\
              &     &     \\
\hline
    +         &     &     \\
 R-Values  & $\chi^2$/DOF $=$ $13.2/17$ & $\chi^2$/DOF $=$ $7.6/17$ \\
 Mult+Sub      & (CL $=$ $72$\%) & (CL $=$ $97$\%) \\
              &     &     \\
\hline
    +         &     &     \\
 Upward-Muons  & $\chi^2$/DOF $=$ $13.6/18$ & $\chi^2$/DOF $=$ $7.7/18$ \\
  Stop/Thru    & (CL $=$ $75$\%) & (CL $=$ $98$\%) \\
               &     &     \\
\hline
    +         &     &     \\
  CHOOZ  & $\chi^2$/DOF $=$ $25.4/28$ & $\chi^2$/DOF $=$ $18.7/28$ \\
  $P(\bar{e} \rightarrow \bar{e})$
                          &  (CL $=$ $61$\%)   &  (CL $=$ $91$\%)   \\
              &     &     \\
\hline
\hline
\end{tabular}
\end{center}
\vspace{5mm}
\noindent Table~1: 
Cumulative breakdown of $\chi^2$ contributions
and confidence levels 
for the various data-sets
(see text) calculated for threefold maximal mixing
with $\Delta m^2 = 0.98 \times 10^{-3}$ eV$^2$
and twofold maximal $\nu_{\mu}-\nu_{\tau}$
mixing with $\Delta m^2 = 2.2 \times 10^{-3}$ eV$^2$.
The $\Delta m^2$ values 
quoted correspond
to the overall $\chi^2$-minimum in each case 
summed over all the data-sets listed.
Both threefold maximal mixing
and twofold maximal $\nu_{\mu}-\nu_{\tau}$
mixing give excellent overall fits
to the data.

\newpage

\noindent {\bf {\large Figure Captions}}

\vspace{10mm}
\noindent Figure~1.
a) The combined KAMIOKA/SUPER-K multi-GeV
zenith angle distribution for a) $\mu$-like events
and b) $e$-like events.
The solid curve in each case
is the threefold maximal mixing
prediction for $\Delta m^2 = 0.98 \times 10^{-3}$ eV$^2$,
including terrestrial matter effects
and compensation effects.
The twofold maximal $\nu_{\mu}-\nu_{\tau}$  mixing
predictions (dashed curve) and 
the threefold vacuum predictions (dotted curve)
are also shown.
Either threefold maximal mixing
with matter effects 
or twofold maximal $\nu_{\mu}-\nu_{\tau}$ mixing 
are consistent with the 
multi-GeV zenith-angle data.

\vspace{10mm}
\noindent Figure~2.
The combined KAMIOKA/SUPER-K data
on the up/down ratios
for $e$-like and $\mu$-like events 
with $|\cos \theta | > 0.2$
(data-points and shaded bands).
The various theoretical predictions
are plotted versus $\Delta m^2$.
Threefold maximal mixing with matter effects
(solid curves)
approaches two-fold maximal 
$\nu_{\mu}-\nu_{\tau}$ mixing
(dashed curves) 
as $\Delta m^2 \rightarrow 0$,
and over-compensation
of the $\nu_e$-rate 
(expected in vacuum, dotted curve)
is suppressed, 
consistent with the data.

\vspace{10mm}
\noindent Figure~3.
The SUPER-K sub-GeV 
zenith angle distributions
(lepton momentum $p > 400$ MeV)
for a) $\mu$-like and b) $e$-like events.
The curves correspond
to threefold maximal mixing
with terrestrial matter effects
and compensation effects
for $\Delta m^2 = 0.98 \times 10^{-3}$ eV$^2$.
The dashed curve
shows the expected
energy-independent 
`matter oscillations'
of amplitude $\pm 1/6$
and density-dependent wavelength,
which are obscured by
angular smearing (solid curve)
in the water-Cherenkov
experiments.

\vspace{10mm}
\noindent Figure~4.
The atmospheric neutrino ratio 
$R = (\mu/e)_{DATA}/(\mu/e)_{MC}$
as measured by SUPER-K
(data-points and shaded bands).
Within the statistical and systematic
errors shown the data for $R$
are consistent with
$\Delta m^2 \sim 10^{-3}$ eV$^2$.  
Threefold maximal mixing
including terrestrial matter effects
(solid curves)
predicts less change in $<R>$
from sub-GeV to multi-GeV energies
(open and filled data points respectively)
than twofold maximal 
$\nu_{\mu}-\nu_{\tau}$ mixing
(dashed curves).

\newpage

\vspace{10mm}
\noindent Figure~5.
The double ratio $S$
of stopping to through muons
in data and monte-carlo in SUPER-K
(data point and shaded band).
Threefold maximal mixing
with matter effects included
(solid curve)
predicts a somewhat lower minimal
suppression, but is otherwise
very similar to the
twofold maximal $\nu_{\mu}-\nu_{\tau}$ mixing
prediction (dashed curve),
with the shift of a factor
$2/3$ on the $\Delta m^2$-scale, 
see Eq.~(14).
The data for $S$ point to
$\Delta m^2 \sim 10^{-3}$ eV$^2$
in both twofold and threefold mixing.

\vspace{10mm}
\noindent Figure~6.
The data from the CHOOZ reactor experiment 
on the $\bar{\nu}_e$ survival probability
$P(\bar{e} \rightarrow \bar{e})$ plotted
as a function of neutrino energy $E = 1.8+E_e$
where $E_e$ is the measured positron energy.
The solid curve shows 
the threefold maximal mixing prediction
for $\Delta m^2 = 0.98 \times 10^{-3}$ eV$^2$.
With continued running,
increased detector mass
and particularly with increased
baseline $L \sim 5-10$ km,
significant
$\bar{\nu}_e$ disappearance
might well be established
in reactor experiments.

\vspace{10mm}
\noindent Figure~7.
The overall $\chi^2$
in our combined fit,
plotted versus $\Delta m^2$.
In twofold maximal $\nu_{\mu}-\nu_{\tau}$ mixing (dashed curve)
there is a broad minimum extending over the range 
$\Delta m^2 \sim 7 \times 10^{-4} - \; 4 \times 10^{-3}$ eV$^2$
with a best-fit
$\Delta m^2 \simeq 2.2 \times 10^{-3}$ eV$^2$
corresponding to an excellent fit
($\chi^2/$DOF $= 18.7/28$, CL $=91$\%).
In threefold maximal mixing (solid curve)
$\Delta m^2$ is relatively
precisely determined:
$\Delta m^2 \simeq (0.98 \pm 
\raisebox{0.9ex}{0.30} \raisebox{-0.9ex}{\hspace{-7.3mm}0.23}
\hspace{0.4mm}) \times 10^{-3}$ eV$^2$,
but still corresponds to a very good fit
($\chi^2/$DOF $= 25.4/28$, CL $=61$\%).

\vspace{10mm}
\noindent Figure~8.
Suppression of $\nu_e$ mixing
by terrestrial matter effects.
The predicted $\nu_e$ appearance probability
as a function of the propagation length $L$
through the Earth,
in threefold maximal mixing with
$\Delta m^2 \equiv 10^{-3}$ eV$^2$.
The dotted curve is for vacuum mixing only
and the solid curve includes matter effects
(for visual clarity
the curves plotted refer to a $\; 50:50 \;$
average over $\nu$/$\bar{\nu}$ beams).
Matter oscillations,
with reduced wavelength $(\sqrt{2}GN_e)^{-1}$,
are evident at large $L$,
but proposed long-baseline accelerator
experiments with $L$ $\simlt$ $1000$ km
are largely unaffected.

\newpage
\pagestyle{empty}
\begin{figure*}[hbt]
\epsfig{figure=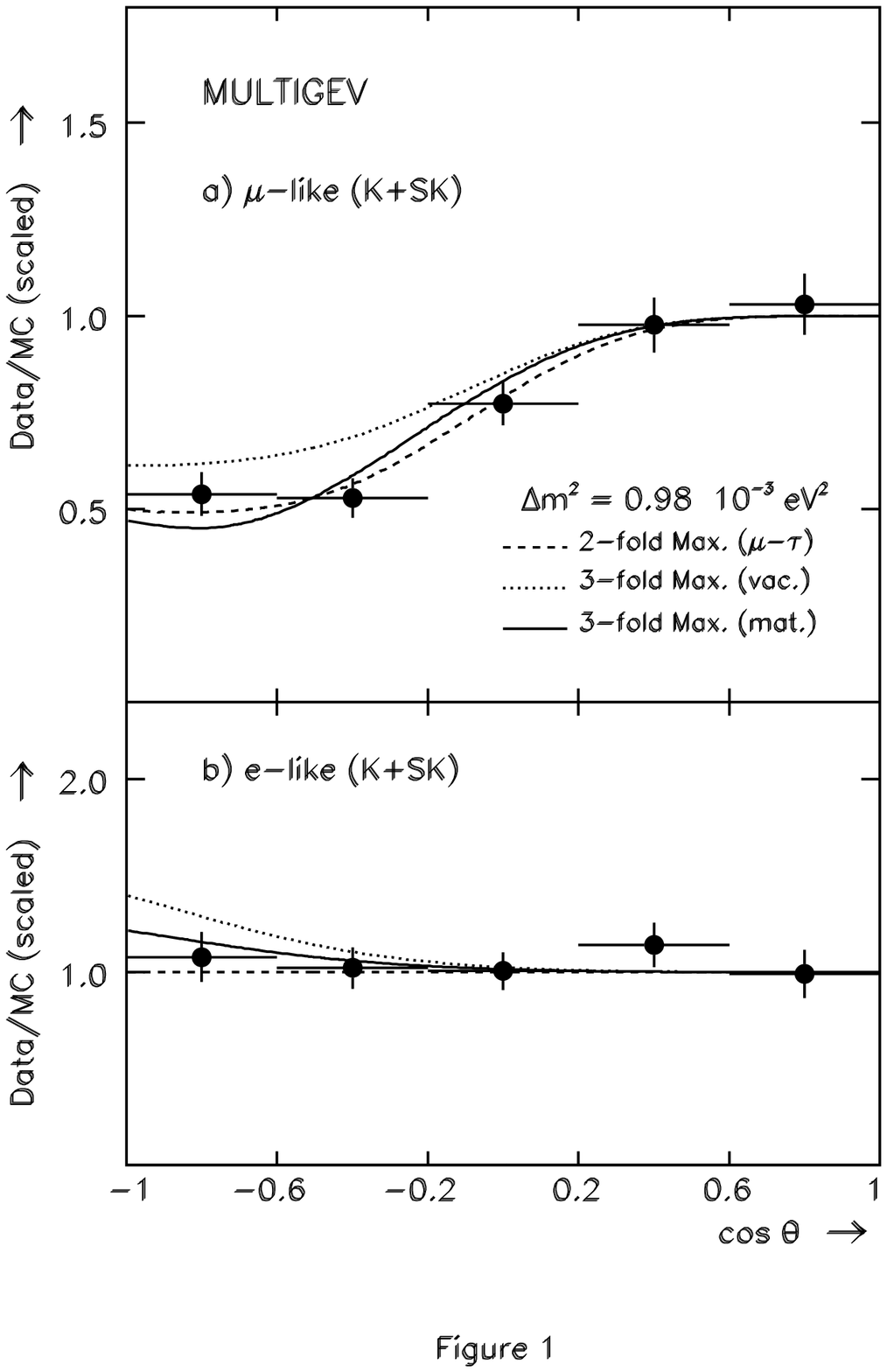,width=150mm,bbllx=100pt,bblly=120pt
,bburx=550pt,bbury=720pt}
\end{figure*}
\newpage
\begin{figure*}[hbt]
\epsfig{figure=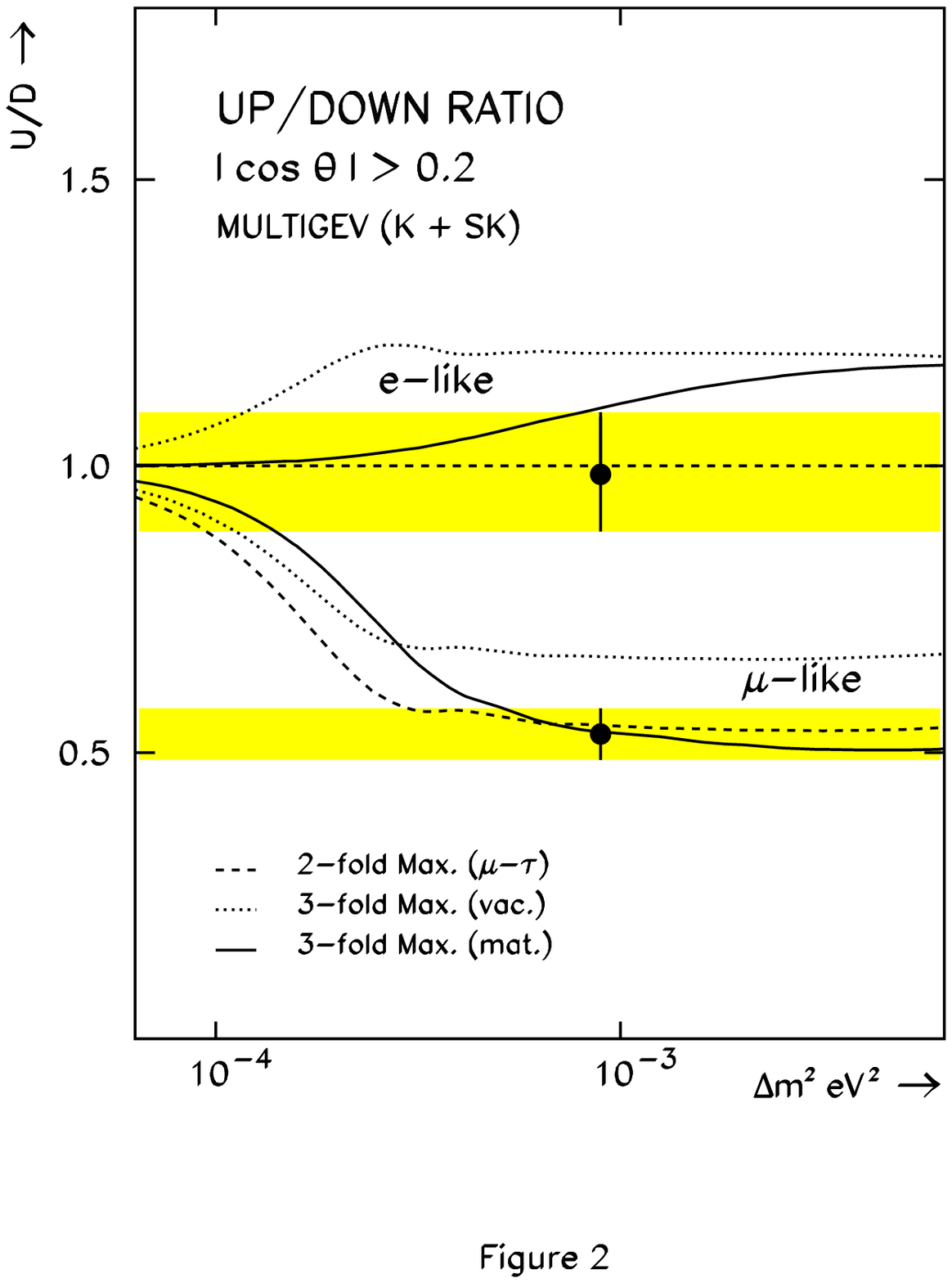,width=150mm,bbllx=100pt,bblly=120pt
,bburx=550pt,bbury=720pt}
\end{figure*}
\newpage
\begin{figure*}[hbt]
\epsfig{figure=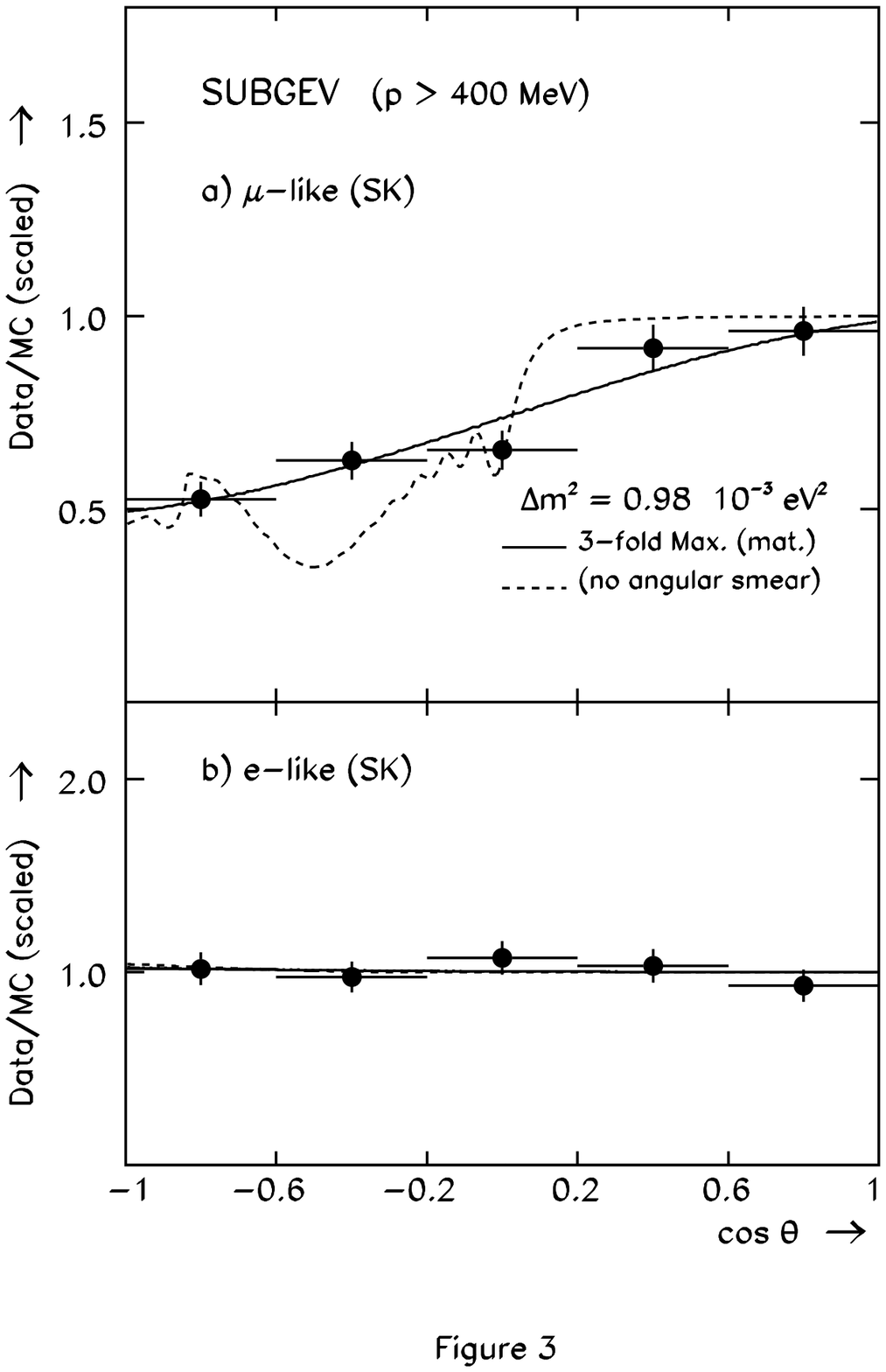,width=150mm,bbllx=80pt,bblly=100pt
,bburx=530pt,bbury=700pt}
\end{figure*}
\newpage
\begin{figure*}[hbt]
\epsfig{figure=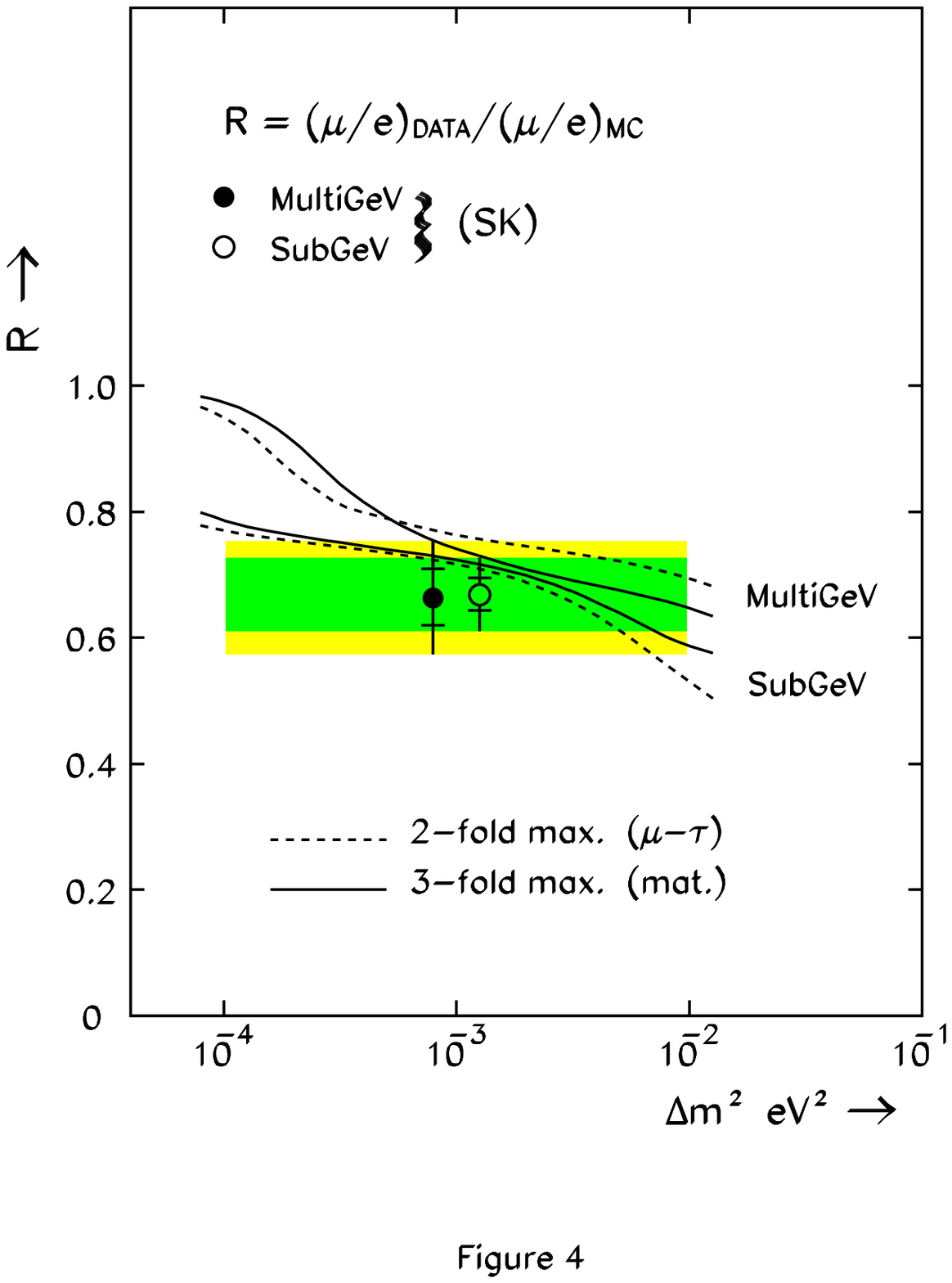,width=150mm,bbllx=80pt,bblly=100pt
,bburx=530pt,bbury=700pt}
\end{figure*}
\newpage
\begin{figure*}[hbt]
\epsfig{figure=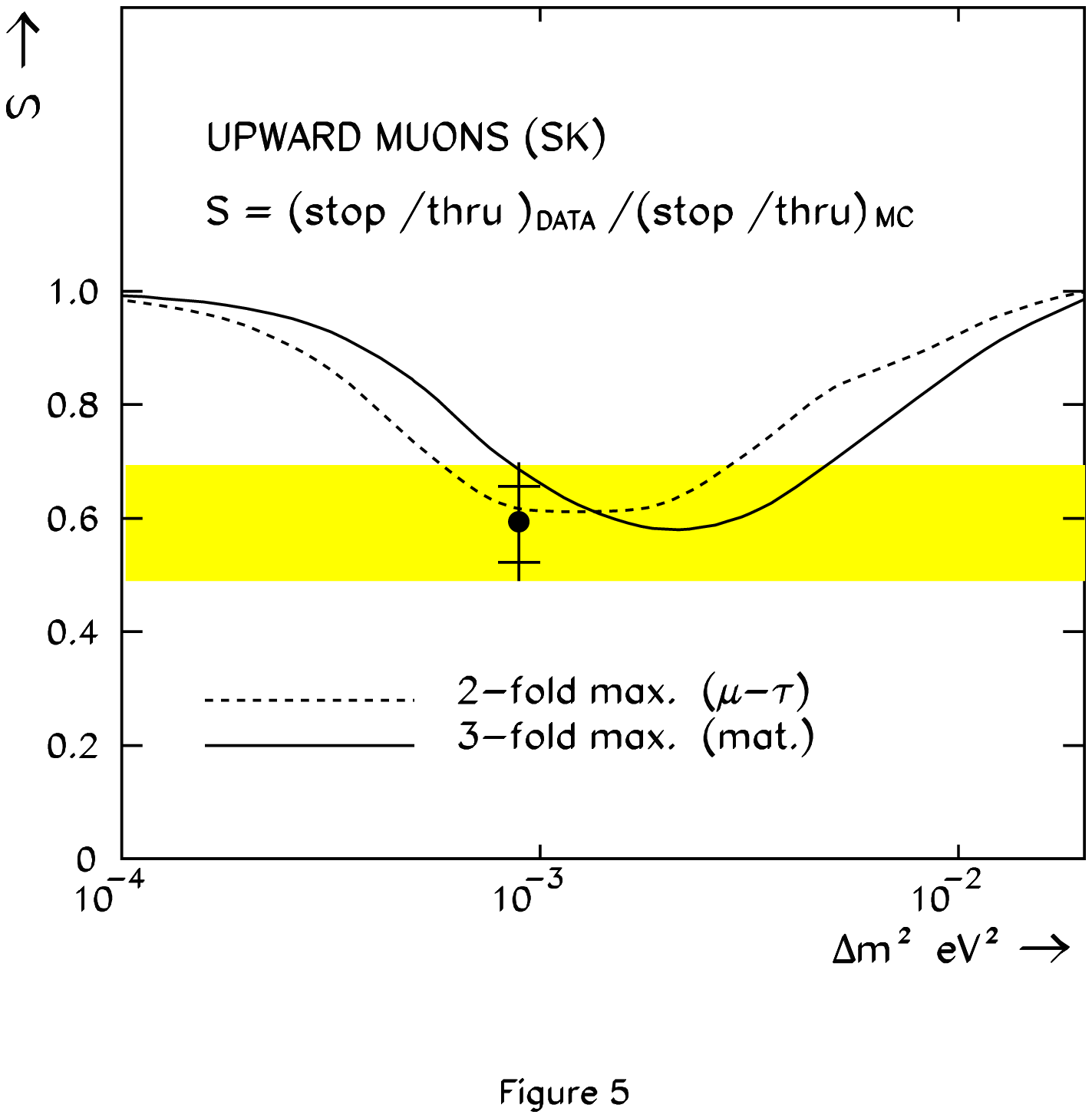,width=150mm,bbllx=80pt,bblly=100pt
,bburx=530pt,bbury=700pt}
\end{figure*}
\newpage
\begin{figure*}[hbt]
\epsfig{figure=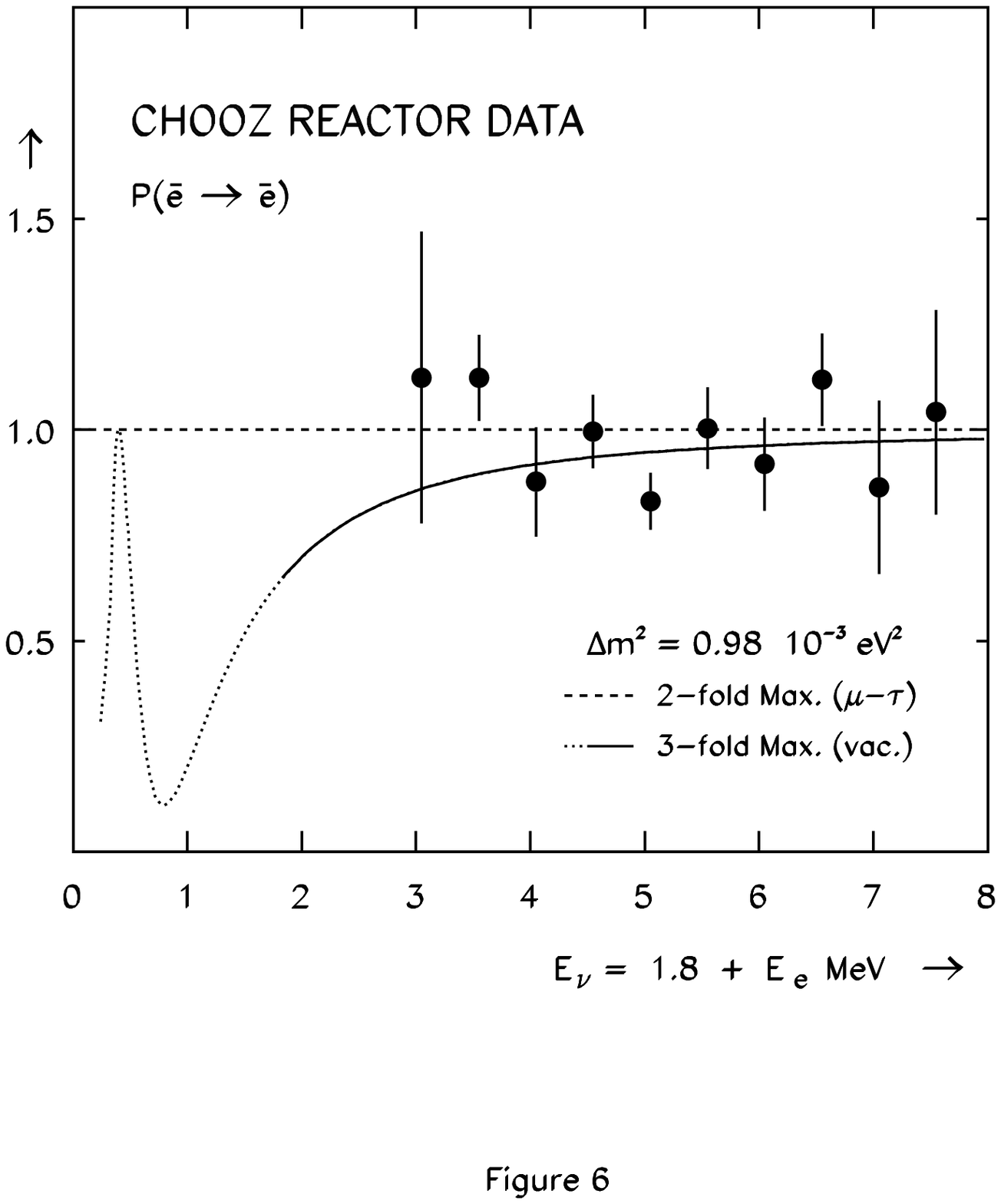,width=150mm,bbllx=80pt,bblly=100pt
,bburx=530pt,bbury=700pt}
\end{figure*}
\newpage
\begin{figure*}[hbt]
\epsfig{figure=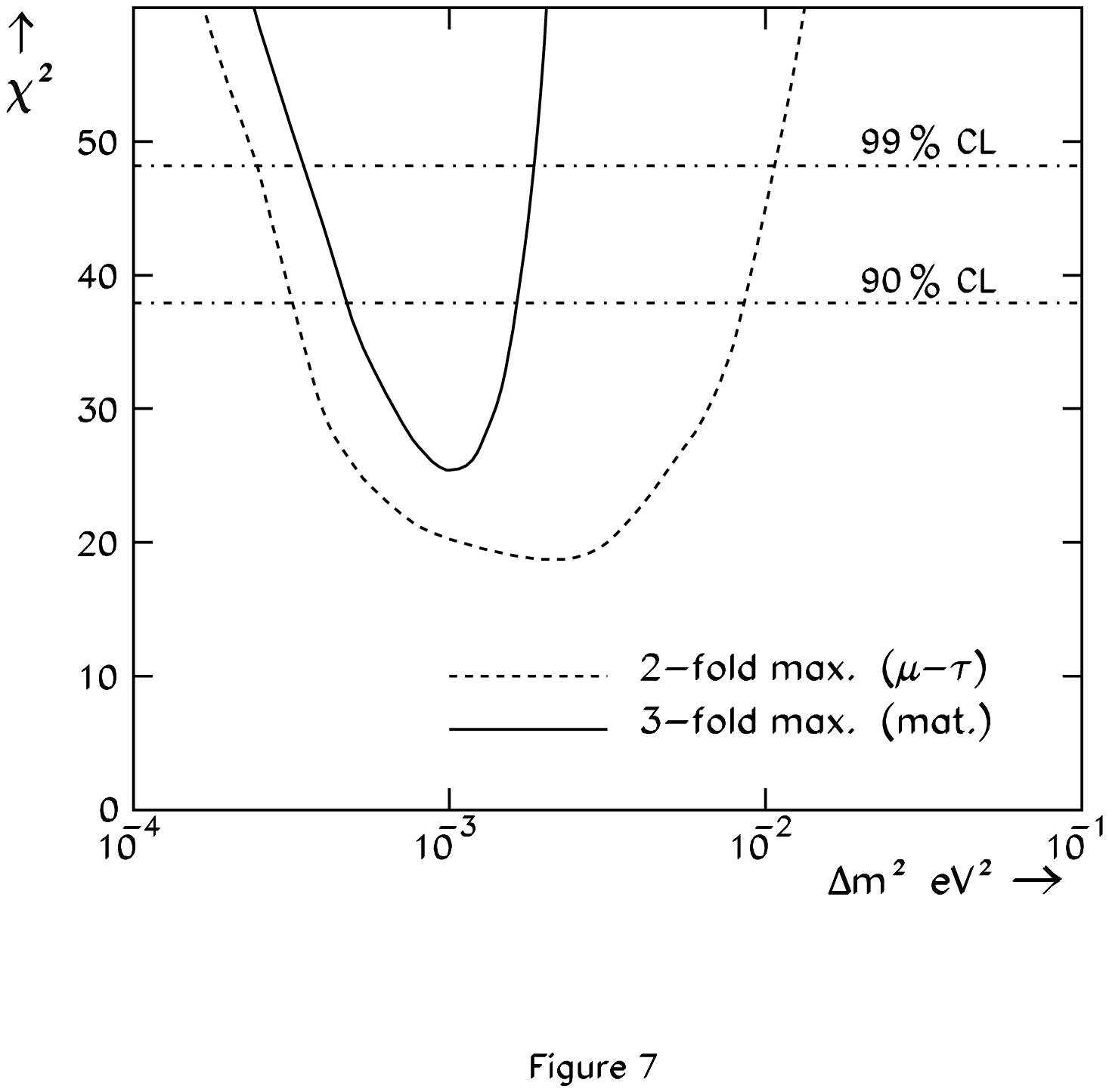,width=150mm,bbllx=80pt,bblly=100pt
,bburx=530pt,bbury=700pt}
\end{figure*}
\newpage
\begin{figure*}[hbt]
\epsfig{figure=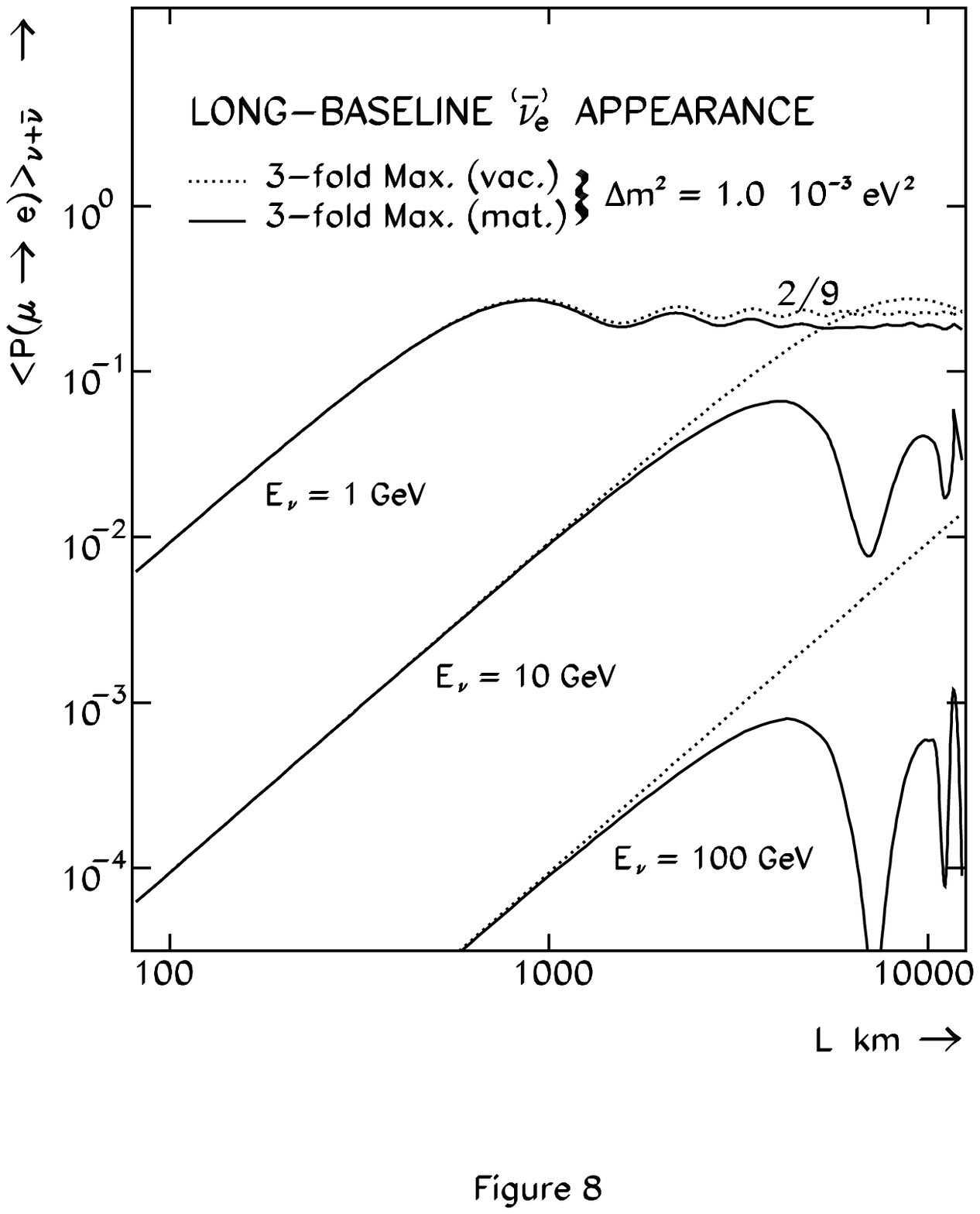,width=150mm,bbllx=80pt,bblly=100pt
,bburx=530pt,bbury=700pt}
\end{figure*}

\newpage 

\begin{thebibliography}{99}
\bibitem[1]{chz1} M. Apollonio et al. Phys. Lett. B 420 (1998) 397. 
                   (hep-ex/9711002).
\bibitem[2]{hps1} P.\ F.\ Harrison, D.\ H.\ Perkins and W.\ G.\ Scott.
                  Phys.\ Lett.\ B 349 (1995) 357.
\bibitem[3]{skam} Y. Fukuda et al. Phys.\ Lett.\ 436 (1998) 33 
                  (hep-ex/9805006). \\
                  Y. Fukuda et al. Phys.\ Lett.\ 433 (1998) 9 
                  (hep-ex/9803006).
\bibitem[4]{kam1} K.\ S.\ Hirata et al. Phys. Lett. B 205 (1988) 416; 
                  B280 (1992) 146. \\
                  Y.\ Suzuki. ICHEP '96. 
                  ed.\ Z. Ajduk and A.\ K.\ Wroblewski. (1997).
\bibitem[5]{skos} Y. Fukuda et al. Phys.\ Rev.\ Lett.\ 81 (1998) 1562 
                  (hep-ex/9807003). \\
                  M. Messier. APS
                  Division of Particles and Fields,
                  UCLA (1999).
\bibitem[6]{kama} Y. Fukuda et al. Phys. Lett. B335 (1994) 237.
\bibitem[7]{ftmt} R.\ Foot, R.\ R.\ Volkas and O.\ Yasuda. 
                  Phys.\ Lett.\ B 433 (1998) 82.
\bibitem[8]{pant} V.\ Barger et al. Phys.\ Rev.\ D22 (1980) 2718. \\
                  J.\ Pantaleone. Phys.\ Rev.\ D49 (1994) 2152;
                  hep-ph/9810467. \\
                  G.\ L.\ Fogli et al.
                  Phys.\ Rev.\ D 59 (1999) 033.001.
\bibitem[9]{hps3} P.\ F.\ Harrison, D.\ H.\ Perkins and W.\ G.\ Scott.
                  Phys.\ Lett.\ B 396 (1997) 186.
\bibitem[10]{wgs1} W.\ G.\ Scott. Nucl.\ Phys.\ B 
                  (Proc.\ Suppl.\ ) 66 (1998) 411.
\bibitem[11]{ster} D.\ O.\ Caldwell and R.\ N.\ Mohapatra.
                    Phys.\ Rev.\ D 48 (1993) 3269. \\
                  J.\ Peltoniemi and J.\ W.\ Valle.
                  Nucl.\ Phys.\ B 408 (1993) 406.
\bibitem[12]{hps2} P.\ F.\ Harrison, D.\ H.\ Perkins and W.\ G.\ Scott.
                  Phys.\ Lett.\ B 374 (1996) 111.
\bibitem[13]{sksn} Y. Fukuda et al.\
                  Phys.\ Rev.\ Lett.\ 81 (1998) 1158;4279
                  (hep-ex/9805021). 
\bibitem[14]{home} B.\ T.\ Cleveland et al.\ 
                   Nucl.\ Phys.\ B (Proc. Suppl.) 38 (1995) 47. 
\bibitem[15]{gnmt} C.\ Giunti, C.\ W.\ Kim and M. Monteno.
                   Nucl.\ Phys.\ B521 (1998) 3.  
\bibitem[16]{jaco} J. A. Jacobs. The Earth's core.  Academic Press (1987). \\
                   A.\ M.\ Dziewonski and D.\ L.\ Anderson.
                   Phys.\ Earth Planet Int.\ 25 (1981) 297.
\bibitem[17]{wolf} L.\ Wolfenstein. Phys.\ Rev.\ D17 (1978) 2369; 
                   D20 (1979) 2634. \\
                   R.\ R.\ Lewis. Phys.\ Rev.\ D21 (1980) 663. \\
                   P.\ Langacker. et al. Phys.\ Rev.\ D27 (1983) 1228.
\bibitem[18]{dhp1} D.\ H.\ Perkins. Nucl.\ Phys.\ B399 (1993) 3.
\bibitem[19]{naka} M.\ Nakahata et al.\ 
                   Nucl.\ Instrum.\ Methods.\ A421 (1999) 113.
                   hep-ex/980727.
\bibitem[20]{evmu} S.\ Kasuga et al.\ Phys.\ Lett.\ B 374 (1996) 238.
\bibitem[21]{phd1} T. Eichten et al.\ Phys.\ Lett.\ B 40 (1973) 281. \\
                   W.\ G.\ Scott. D.\ Phil.\ Thesis. Oxford (1975).
\bibitem[22]{ftud} R.\ Foot, R.\ R.\ Volkas and O.\ Yasuda. 
                   Phys.\ Lett.\ B 421 (1998) 245.
\bibitem[23]{bart} P.\ Lipari, T.\ Stanev and T.\ K.\ Gaisser.
                   Phys.\ Rev.\ D 58 (1998) 073003. \\  
                   T.\ K.\ Gaisser. New Era in Neutrino Physics. Tokyo (1998)
                   hep-ph/9811314.
\bibitem[24]{lsco} J.\ M.\ LoSecco. hep-ph/9807359.
\bibitem[25]{maco} M. Ambrosio et al. Phys.\ Lett.\ B434 (1998) 451
                   (hep-ex/9807005).
\bibitem[26]{kaum} T.\ Hatakeyama et al.
                   Phys.\ Rev.\ Lett.\ 81 (1998) 2016
                   (hep-ex/9806038). 
\bibitem[27]{skum} Y.\ Fukuda et al.\  hep-ex/9812014.
                   T.\ Kajita. hep-ex/9810001. \\
                   Y.\ Suzuki. 
                   WIN99 Capetown (1999).
\bibitem[28]{palo} F. Boehm.
                   VIII International Workshop on Neutrino Telescopes.
                   Venice. (1999).
\bibitem[29]{kmln}  T. Alivisatos et al.\ 
                    Stanford-HEP-98-03;Tohoku-RCNS-98-15. (1998).
\bibitem[30]{xing} H.\ Fritzsch and Z.\ Xing. 
                   Phys.\ lett.\ B 372 (1996) 265; B 440 (1998) 313.  
\bibitem[31]{bimx} V.\ Barger et al.
                   Phys.\ Lett.\ B437 (1998) 107. \\ 
                   A.\ J.\ Bahz, A.\ S.\ Goldhaber and M.\ Goldhaber.
                   Phys.\ Rev.\ Lett.\ 81 (1998) 5730. \\
                   D.\ V.\ Ahluwalla. Mod.\ Phys.\ Lett.\ A 18 (1998) 2249. \\
                   H.\ Giorgi and S.\ L.\ Glashow. hep-ph/9808293. \\
                   W.\ G.\ Scott. IDM98 Buxton (1998). RAL-TR-1998-072.
\bibitem[32]{alt1} G.\ Altarelli and F.\ Feruglo. 
                   Phys.\ Lett. B439 (1998) 112; JHEP 9811:021 (1998).
\bibitem[33]{jarl} R.\ N.\ Mohapatra and S.\ Nussinov. hep-ph/9809415. \\
                   C.\ Giunti. hep-ph/9810272. \\
                   C.\ Jarlskog et al.\ hep-ph/9812282. \\
                   R.\ Barbieri et al.\ hep-ph/9901228. \\
                   S.\ Lola and G.\ G.\ Ross. hep-ph/9902283.
\bibitem[34]{lsnd} C. Athanassopoulos et al.\ 
                  Phys.\ Rev.\ Lett.\ 75 (1995) 2650; 81 (1998) 1774. \\
                  Phys. Rev. C 54 (1996) 2685.

\end{thebibliography}
\end{document}